\documentclass[a4paper,english,aps,pre,showpacs]{revtex4}
\usepackage[T1]{fontenc}
\usepackage[latin1]{inputenc}
\usepackage{amsmath}
\usepackage{amssymb}

\makeatletter

\providecommand{\LyX}{L\kern-.1667em\lower.25em\hbox{Y}\kern-.125emX\@}

\usepackage{babel}
\makeatother
\begin{document}
\pacs{02.70.Hm, 61.43.Fs, 02.30.Nw, 61.43.Bn}

\title{Instantaneous frequency and amplitude identification using wavelets:
Application to glass structure}

\author{J.~D.~Harrop, S.~N.~Taraskin and S.~R.~Elliott}

\email{jdh30@cam.ac.uk}

\homepage{http://www.chem.pwf.cam.ac.uk/~jdh30}

\affiliation{Department of Chemistry, University of Cambridge, Lensfield Road,
Cambridge CB2 1EW, United Kingdom.}

\maketitle
We regret that it has come to our notice that some typographical errors
appeared in the paper \cite{Harrop2002} in the description of the
new mother wavelet function and associated normalization expressions.

The correct expressions for the mother wavelet function are (the equation
numbers refer to those in the original paper):\[
\psi (t;\sigma )=e^{-t^{2}/2}\left\{ p(\sigma )\left[\cos (\sigma t)-\kappa (\sigma )\right]+iq(\sigma )\sin (\sigma t)\right\} \tag {5}\]
\begin{align*}
p(\sigma ) & = \pi ^{-1/4}(1+3e^{-\sigma ^{2}}-4e^{-3\sigma ^{2}/4})^{-1/2} \tag{6a} \\
q(\sigma ) & = \pi ^{-1/4}(1-e^{-\sigma ^{2}})^{-1/2}. \tag{6b}
\end{align*}

The variance of the Gaussian approximation to the envelope of the
mother wavelet is:\[
\sigma _{\psi }^{2}(\sigma )=\frac{1}{4}\sqrt{\pi }\left\{ q^{2}\left[1+\left(2\sigma ^{2}-1\right)e^{-\sigma ^{2}}\right]+p^{2}\left\{ \left[1+\left(3-2\sigma ^{2}\right)e^{-\sigma ^{2}}\right]-2\left(2-\sigma ^{2}\right)e^{-3\sigma ^{2}/4}\right\} \right\} .\tag {11}\]

With this formulation, the remainder of the paper is correct and all
results are unaffected.

\begin{acknowledgments}
We are grateful to Victor~P.~Ostanin for pointing out the mistake.
\end{acknowledgments}

\bibliographystyle{revtex}

\begin{thebibliography}{1}
\expandafter\ifx\csname natexlab\endcsname\relax\def\natexlab#1{#1}\fi
\expandafter\ifx\csname bibnamefont\endcsname\relax
  \def\bibnamefont#1{#1}\fi
\expandafter\ifx\csname bibfnamefont\endcsname\relax
  \def\bibfnamefont#1{#1}\fi
\expandafter\ifx\csname citenamefont\endcsname\relax
  \def\citenamefont#1{#1}\fi
\expandafter\ifx\csname url\endcsname\relax
  \def\url#1{\texttt{#1}}\fi
\expandafter\ifx\csname urlprefix\endcsname\relax\def\urlprefix{URL }\fi
\providecommand{\bibinfo}[2]{#2}
\providecommand{\eprint}[2][]{\url{#2}}

\bibitem[{\citenamefont{Harrop et~al.}(2002)\citenamefont{Harrop, Taraskin, and
  Elliott}}]{Harrop2002}
\bibinfo{author}{\bibfnamefont{J.~D.} \bibnamefont{Harrop}},
  \bibinfo{author}{\bibfnamefont{S.~N.} \bibnamefont{Taraskin}},
  \bibnamefont{and} \bibinfo{author}{\bibfnamefont{S.~R.}
  \bibnamefont{Elliott}}, \bibinfo{journal}{Phys. Rev. E}
  \textbf{\bibinfo{volume}{66}}, \bibinfo{pages}{026703}
  (\bibinfo{year}{2002}).

\end{thebibliography}

\end{document}